\documentclass[a4paper,12pt,english]{article}
\usepackage{amssymb}
\usepackage{amsfonts}
\usepackage{graphicx}
\usepackage{amsmath}
\usepackage{color}

\newcommand{\be}{\begin{equation}}
\newcommand{\ee}{\end{equation}}
\begin{document}


\begin{titlepage}
\begin{center}

\noindent{{\LARGE{On non-homogeneous tachyon condensation in closed string theory}}}

\smallskip
\smallskip

\smallskip
\smallskip
\smallskip
\smallskip
\noindent{\large{Gaston Giribet$^{1,2}$, Laura Rado$^{3}$}}

\smallskip
\smallskip

\end{center}

\smallskip
\smallskip
\centerline{$^1$ Departamento de F\'{\i}sica, Universidad de Buenos Aires and IFIBA-CONICET}
\centerline{{\it Ciudad Universitaria, Pabell\'on 1, 1428, Buenos Aires, Argentina.}}

\smallskip
\smallskip
\centerline{$^2$ Martin Fisher School of Physics, Brandeis University}
\centerline{{\it Waltham, Massachusetts 02453 USA.}}

\smallskip
\smallskip

\centerline{$^3$ Instituto de F\'{\i}sica, Universidade de S\~{a}o Paulo}
\centerline{{\it Rua do Mat\~{a}o Travessa 1371, 05508-090 S\~{a}o Paulo, SP. Brazil}}

\bigskip

\bigskip

\bigskip

\bigskip

\begin{abstract}
Lorentzian continuation of the Sine-Liouville model describes non-homogeneous rolling closed string tachyon. Via T-duality, this relates to the gauged $H_+^3$ Wess-Zumino-Witten model at subcritical level. This model is exactly solvable. We give a closed formula for the $3$-point correlation functions for the model at level $k $ within the range $0<k<2$, which relates to the analogous quantity for $k>2$ in a similar way as how the Harlow-Maltz-Witten $3$-point function of timelike Liouville field theory relates to the analytic continuation of the Dorn-Otto-Zamolodchikov-Zamolodchikov structure constants: We find that the ratio between both $3$-point functions can be written in terms of quotients of Jacobi's $\theta $-functions, while their product exhibits remarkable cancellations and eventually factorizes. Our formula is consistent with previous proposals made in the literature.
\end{abstract}

\end{titlepage}

\section{Introduction}

Tachyon condensation in string theory is a long-standing problem, as it is a phenomenon rather difficult to apprehend both from the conceptual and from the computational points of view. On the one hand, this phenomenon has drastic effects on the spacetime itself, which makes the whole picture difficult to capture. On the other hand, this demands to deal with exact, time-dependent solutions of string theory, which are available only in a few special cases.

The physics of tachyon condensation has been first and better understood in the context of open string theory \cite{Sen, Sen2, Sen3, Sen4}, as well as in some scenarios involving localized closed string tachyons \cite{Polchinski, Minwalla,	Martinec, Cunha}. In the case of the bosonic open string, a method that resulted fruitful to address the problem was resorting to exact conformal field theories (CFT) that admit the interpretation of a string $\sigma $-model on a tachyonic background. In the case of homogeneous tachyonic background, this is typically given by a timelike version of Liouville field theory coupled to $c=1$ matter; that is, by the theory governed by the action
\begin{equation}
S=\frac{1}{8\pi } \int d^2z \Big( \partial X^1 \bar\partial X^1 + \partial \varphi \bar\partial\varphi +({ b+1/b })R\varphi+ 8\pi \mu \  e^{\sqrt{2}b\varphi } \Big) \label{LKJH}
\end{equation}
analytically continued to $\varphi \to X^0=-i\varphi $, $b\to \hat{b}=ib$, which has central charge $c=2-6(\hat{b}-1/\hat{b})^2$.

For open strings, such CFT description in terms of timelike Liouville theory with boundaries has been studied in references \cite{Strominger, Gutperle, Larsen, Constable}; see references therein and thereof. Apart from this continuous approach, the problem has also been studied with other formalisms, like the matrix model approach \cite{MatrixModel, MatrixModel2}. 

In the case of closed string theory, describing homogeneous tachyon condensation in terms of exact time-dependent solutions of the worldsheet CFT is also feasible, and this has been studied, for instance, in references \cite{StromingerTakayanagi, Schomerus, Freedman}. This approach has made possible to achieve important progress in understanding the physical process, as well as it contributed to get a better understanding of the worldsheet CFT itself, leading to very interesting discussions about which is the adequate procedure to analytically extend the standard spacelike Liouville theory in order to produce its timelike version. These technical discussions were mostly in relation to the 3-point function \cite{StromingerTakayanagi, Schomerus}. More recently, the timelike\footnote{There is an abuse of terminology here: Although the theory is usually referred to as timelike, what is motivated by its Lagrangian representation, it was argued in \cite{Ribault} that a careful analysis of the spectrum reveals that the theory with $c<1$ is actually spacelike. Moreover, the authors of \cite{Ribault} explained that a timelike theory would lead to a ill-defined OPE. This, however, does not prevent one from analytically continuing the theory to $c<1$.} Liouville theory --and in particular its correlation functions-- has been reconsidered in \cite{Ribault, H-M-Witten, Giribet1, Schomerus2012, Picco, McElgin,  timelike, timelike2}.

A worldsheet description is also available in the case of non-homogeneous tachyon condensation, which is the problem we will be concerned with here. This has been originally discussed in \cite{Cunha}, where a CFT worldsheet description was proposed to be given by the analytic continuation of the so-called Sine-Liouville theory; namely by the theory governed by the action  
\begin{equation}
S=\frac{1}{8\pi } \int d^2z \Big( \partial X^1 \bar\partial X^1 + \partial \varphi \bar\partial\varphi +{b}R\varphi + 8\pi \mu \ e^{\frac{1}{\sqrt{2}b}\varphi }\cos (\sqrt{k/2}\tilde{X}^1)\Big) \label{HJKL}
\end{equation}
analytically continued to $\varphi \to X^0=-i\varphi $, $b\equiv 1/\sqrt{k-2}\to \hat{b}=ib$. $\tilde{X}$ here represents the T-dual direction associated to $X$. The theory defined in this way has central charge $c=2-6\hat{b}^{-2}$.

The interpretation of the timelike version of the model (\ref{HJKL}) as describing a non-homogeneous rolling tachyon of bosonic closed string theory has been reconsidered in \cite{Yasuaki2}, where the model was studied in terms of its T-dual counterpart, the gauged $H^+_3=SL(2, \mathbb{C})/SU(2)$ Wess-Zumino-Witten (WZW) theory. 

Sine-Liouville theory (\ref{HJKL}) is dual to the gauged $H^+_3$ WZW theory\footnote{More precisely, the duality holds with the $H^+_3/U(1)$ coset model. However, the relevant part of the functional dependence of the correlators is captured by the $H^+_3$ model.} with level $k=2+b^{-2}$. This is known as the Fateev-Zamolodchikov-Zamolodchikov (FZZ) duality \cite{FZZ, FZZ2, FZZ3}, which is a kind of T-duality \cite{FZZ4}. If one extends the FZZ conjecture to negative values of $b^2$, which correspond to the values for which the timelike version of (\ref{HJKL}) is defined, then one is led to state that non-homogeneous tachyon condensation is governed by the WZW action at subcritical level $k<2$; that is, values of the Kac-Moody level that are below the Coxeter number. In particular, this implies that the problem of computing string observables in the time (and $X$) dependent tachyon background would reduce to the problem of computing correlation functions in the subcritical WZW theory. This represents indeed an advantage because, for reasons we will comment below, computing WZW correlators is simpler than computing them in presence of the Sine-Liouville deformation\footnote{Furthermore, being the consistency of the Lagrangian representation (\ref{HJKL}) in the timelike case unclear, this can be thought of as the very definition of what the timelike theory actually means.}. Then, the idea is simple: extending the WZW 3-point function to $k<2$. The problem is similar to that of extending the Liouville 3-point function to values of the central charge $c<1$; i.e. to solve the timelike model. The timelike Liouville 3-point function was discussed in \cite{Zamolodchikov, Petkova, Petkova2, Petkova3, H-M-Witten, Giribet1}. In particular, in \cite{H-M-Witten} Harlow, Maltz, and Witten showed that a proposal for the timelike 3-point function made by Zamolodchikov in \cite{Zamolodchikov} can also be computed by the original Liouville path integral evaluated on a new integration cycle. In Ref. \cite{Giribet1}, it was shown that the expression found in \cite{H-M-Witten} can also be obtained in the Coulomb gas approach by means of the adequate analytic extension of the Selberg type integrals involved. This reproduces the exact result, including the right normalization. Encouraged by this, in this paper we undertake the type of computation of \cite{Giribet1} in the case of WZW 3-point function. This amounts to adapt the calculations of reference \cite{Becker, GN2, GN3} to the range\footnote{It is probable that our result is still valid for negative level $k$. The reason why we mention the lower bound here responds to our belief that stating the validity for values $k<0$ would demand a better understanding of the limit $k\to 0$, which seems to be peculiar in many respects \cite{Nakayama}. However, it was argued by Sylvain Ribault \cite{RibaultPrivate} that there is no obvious obstruction to consider the result also in the range $-\infty <k <2$; see also \cite{RibaultNuevo}.} $0<k<2$. This, as we will see, yields an unexpected result.  

The main reason why the computation based on the WZW theory is more convenient than the one in Sine-Liouville theory is that, in the former, as it happens in the undeformed Liouville theory, a Coulomb gas realization of the 3-point function amounts to deal with the multiple Selberg type integrals of the class computed by Dotsenko and Fateev in the context of the Minimal Models \cite{Dotsenko}, and it has been understood in \cite{Giribet1} how to extend such formulae to the timelike case. In contrast, the computation in the representation (\ref{HJKL}) requires either generalizations of such integrals that are only known in some special cases \cite{FukudaHosomichi}, or duality relations between different multiple integrals that make the calculation notably more involved \cite{Giribet2017}. Here, we will solve the problem in quite efficient way: In section 2, we will review the free field representation of the $H^+_3$ WZW model and the computation of the correlation functions in the Coulomb gas approach. In section 3, we will perform the analogous computation at the subcritical level $k<2$. This will demand a careful analysis of the analytic extension. In section 4, we will rewrite our result in a convenient way, in terms of Jacobi $\theta _1$-function. This will enable us, in section 5, to perform a comparison with other proposals for the subcritical 3-point function that appeared in the literature. Section 6 contains our conclusions.  

\section{The $H_3^+$ WZW model revisited}

We start by reviewing the $H_3^+$ WZW model in the Coulomb gas approach. The purpose is collecting the main formulae and introduce the ingredients for the computation of the $3$-point function and its extension to values $k<2$. In order to facilitate the comparison, we will follow closely the conventions of Refs. \cite{Becker, GN2, GN3}. We refer to those papers for details. 

In a particular representation, the action of the WZW theory is given by
\begin{eqnarray}\label{SWZNW}
S_{\mu} =\frac{1}{8\pi }\int d^{2}z\left [ \sigma \partial \phi \bar{\partial }\phi-\frac{\sqrt{2}}{\alpha _{+}} R\phi +\beta \bar{\partial } \gamma + \bar{\beta }\partial \bar{\gamma }-8\pi \mu\beta \bar{\beta }e^{-(2/\alpha_{+})\phi}\right ].
\end{eqnarray}
where\footnote{The reason why the level $k$ does not appear as an overall factor here is that the field $\phi $ suffers a renormalization by a factor $\sqrt{k-2}$ and absorb a shifted $k\to k-2$ factor emerging through quantum corrections. See \cite{Satoh} for details.} $\alpha_+ ^2=2\sigma (k-2)$. The introduction of $\sigma =\pm 1$ in the Lagrangian enables us to switch between the standard (spacelike) version of the model, corresponding to $k>2$, and its timelike version, which is equivalent to considering $0<k<2$. The sign of $\sigma $ is such that the dilatonic background charge and the exponential potential remain real for both ranges. In this sense, the transition from $k>2$ to $k<2$ is equivalent to a change in the signature of the field $\phi $, which in the free theory ($\mu = 0$) has the correlator 
\begin{equation} \label{propagatorsPhi}
\left \langle \phi (z)\phi (w) \right \rangle =-\sigma\ln(z-w).
\end{equation}

In other words, the case $k<2$ corresponds to $\sigma =-1$, while the case $k>2$ corresponds to $\sigma=+1$; and the WZW model with $\sigma=-1$ can be obtained from the standard case, $\sigma=+1$, by going to imaginary values of the background charge $\alpha_{+}\to i\alpha_{+}$ and, at the same time, Wick rotating the field as $\phi\rightarrow i\phi$. The theory also involves a $\beta -\gamma $ ghost system, which is unaffected by the Wick rotation and in the free theory yields
\begin{equation} \label{propagators}
\left \langle \gamma (z)\beta (w) \right \rangle =-\frac{1}{(z-w)}.
\end{equation}

The holomorphic component of the stress tensor of the theory is given by
\begin{eqnarray}
T(z)=-\frac{1}{2}\sigma \partial\phi (z)\partial\phi (z)-\frac{1}{\alpha_+}\partial^2\phi (z)+\beta(z) \partial\gamma(z) ,  \label{stresstensor}
\end{eqnarray}
and analogously for its complex conjugate counterpart $\bar{T}(\bar{z})$. 

The vertex operators, $V_{j,m,\bar{m}} (z,\bar{z})$, are in correspondence with Kac-Moody primary states $\vert j,m,\bar{m} \rangle = \lim_{z\to \infty} V_{j,m,\bar{m}} (z,\bar{z})\vert 0 \rangle$, which are labeled by $SL(2,\mathbb{R})\times SL(2,\mathbb{R})$ unitary representations\footnote{In defining the $H_3^+$ model, the isospin index $j$ takes the values of the continuous principal series, $j=-1/2+i\lambda $ with $\lambda \in \mathbb{R}$ and $m\in \mathbb{R}$.}. These operators are Virasoro primaries with respect to (\ref{stresstensor}), and are essential elements of the theory. We will be interested in computing their correlation functions 
\begin{eqnarray}\label{A_n}
\mathcal{A}_{m_1,m_2,...m_n}^{j_1,j_2,...j_n}=\Big\langle \prod_{i=1}^{n} V_{j_{i},m_{i},\bar{m}_{i}} (z_{i},\bar{z}_{i}) \Big\rangle _{S_{\mu }} = \int \mathcal{D}\gamma \mathcal{D}\beta \mathcal{D}\phi \ e^{-S_{\mu}}  \prod_{i=1}^{n} V_{j_{i},m_{i},\bar{m}_{i}} (z_{i},\bar{z}_{i}), \label{Antes}
\end{eqnarray}
defined with the action (\ref{SWZNW}). A suitable representation of the vertex operators is given by
\begin{eqnarray}\label{vertex}
V_{j,m,\bar{m}}(z,\bar{z})=\gamma(z)^{j-m}\bar{\gamma}(\bar{z})^{j-\bar{m}}e^{\frac{2}{\alpha_{+}}j\phi(z,\bar{z}) }.
\end{eqnarray}

In the Coulomb gas formalism, the $n$-point correlation functions (\ref{Antes}) are defined by adding, apart from the $n$ operators (\ref{vertex}), extra operators that play the role of screening the background charge in (\ref{SWZNW}). These screening operators are given by
\begin{eqnarray}\label{screening S+}
\mathcal{S}_1=\int d^2 w\beta(w)\bar{\beta}(\bar{w})e^{-\frac{2}{\alpha _{+}}\phi(w,\bar{w})},
\end{eqnarray}
and
\begin{eqnarray}\label{screening S-}
\mathcal{S}_2=\int d^2 w(\beta(w)\bar{\beta}(\bar{w}))^{(k-2)}e^{-\sigma(\alpha _{+} \phi(w,\bar{w}))}.
\end{eqnarray}
These are marginal operators, meaning that they commute with the Kac-Moody currents that generate the affine symmetry of the theory and have conformal dimension (1,1) with respect to the stress tensor (\ref{stresstensor}).

There are some advantages in considering $\mathcal{S}_2$ over $\mathcal{S}_1$, and the use of any of them leads to the same result \cite{GN3}. In particular, when one faces the problem of extending the WZW model to $k<2$, the advantage of using the operator $\mathcal{S}_2$ is that the amount of them to be inserted in such case, say $s$, is a function of the states momenta $j_i$ and $k$ that does not depend on whether the level is grater or lower than the critical value $2$. This is given by $s=(1+j_1+j_2+j_3)/(k-2)$.

The situation is similar in Liouville field theory, where there also exist two dimension-one operators that can be used as screening charges. From the instrumental point of view, the use of either (\ref{screening S+}) or (\ref{screening S-}) should merely be regarded as a computational trick since, as said, it has been explicitly shown in \cite{GN3} that the use of any of them eventually leads to the same result. This is a consequence of the weak-strong duality that the theory exhibits under the interchange ${k-2}\leftrightarrow 1/({k-2})$ at quantum level\footnote{Although connected, this strong-weak duality is different from the duality of Liouville theory under $b\leftrightarrow 1/b$, for instance in that the former does not leave the WZW central charge invariant. Still, the observables exhibit such a symmetry, which is usually regarded as Langlands duality.}. From the conceptual point of view, however, the interpretation of the screening operators (\ref{screening S+}) and (\ref{screening S-}) is quite different, being that of the latter more subtle due to its dependence with $k$ in the exponential and the higher-order form its $\beta$-dependent part takes when bosonizing the ghost system. The physical interpretation of the operator (\ref{screening S-}) in the $k>2$ WZW theory has been given in \cite{Ranjbar}, where it was identified as the operator responsible for finite-$k$ effects associated to worldsheet instantons \cite{MO3}. 

These non-perturbative operators (\ref{screening S-}) in the timelike theory happen to scale in the same manner as how the dual cosmological constant operator of Liouville theory, $e^{\sqrt{2}\varphi /b}$, scales in the $c<1$ case: While under the Wick rotation $(b,\varphi )\to (ib,-i\varphi )$ the operator $e^{\sqrt{2}b\varphi }$ transforms into itself, the dual operator $e^{+\sqrt{2}\varphi /b}$ changes to $e^{-\sqrt{2}\varphi /b}$. That is, in the timelike case both operators blow up in opposite directions of the field space. The same occurs in the WZW through the Wick rotation $(\alpha_+ ,\phi )\to(i\alpha_+ ,i\phi )$, where the operator (\ref{screening S-}) takes the form $(\beta\bar{\beta})^{-\alpha_+^2/2}e^{+\alpha_+\phi}$. This sign difference in the exponential with respect to the spacelike case is relevant for the spacetime interpretation and for the validity of the perturbation theory. Typically, the presence of the Liouville type potential barriers prevents the strings to explore the zone of strong-coupling, where the linear dilaton grows dangerously. Unlike what happens in the theory with $k>2$, in the subcritical WZW theory the operators (\ref{screening S+}) blows up when $\phi \to -\infty $ while (\ref{screening S-}) blows up in the opposite direction $\phi \to +\infty $. It is the operator (\ref{screening S-}) the adequate one to perform the Coulomb gas computation of the $k<2$ correlators perturbatively\footnote{The situation of the two screening operators in the subcritical WZW theory is totally analogous to that of timelike Liouville field theory. In fact, it is worth mentioning that the relation between the coupling constants of operators (\ref{screening S+}) and (\ref{screening S-}) is not arbitrary in the WZW theory, but is given by a formula analogous to the one that relates the standard and the dual cosmological constants in Liouville theory. This has been derived in \cite{GN3}. Therefore, the interesting discussion of \cite{McElgin} about the presence of the dual operator for special values of $b$ applies here as well. We will not repeat the discussion here because we are interested in generic values of the central charge. We rather refer to section 8 of reference \cite{McElgin}.}.

In the Coulomb gas approach, $s$ is the amount of integrals to be solved. Therefore, the resulting expressions in principle only make sense for $s\in \mathbb{Z}_{\geq 0}$, and an analytic extension is needed in order to gather configurations that correspond to other values of $s$. On the other hand, $s$ being an integer number, such analytic extension is not unique. Here, we will give a precise prescription. The standard procedure to deal with this problem, which has been shown to work well in diverse setups, is to first assume kinematic configurations yielding $s\in \mathbb{Z}_{\geq 0}$, perform the integration over the worldsheet variables, and eventually extend the final result to more general values of $s$.

Using the vertex operator (\ref{vertex}) and the screening operator (\ref{screening S-}), for $n=3$ we can write
\begin{eqnarray}\nonumber
&&\mathcal{A}_{j_1,m_2,m_3}^{j_1,j_2,j_3}= \mu^{s_\sigma}\Gamma(-s_\sigma)\int\prod_{t=1}^{s_\sigma} d^2 w_t \left\langle  e^{2j_1\tilde{\phi}(0)/\alpha_{+}}e^{2j_2\tilde{\phi}(1)/\alpha_{+}}e^{2j_3\tilde{\phi}(\infty)/\alpha_{+}} \prod_{t=1}^{s_\sigma}e^{-\sigma(\alpha _{+} \tilde{\phi}(w_t,\bar{w_t}))}\right\rangle _{S_0}\\ 
&&\times \left\langle \gamma ^{j_2-m_2}(1) \gamma ^{j_3-m_3}(\infty)\prod_{t=1}^{s_\sigma}\beta(w_t)^{(k-2)} \right\rangle _{S_0} \left\langle \bar{\gamma }^{j_2-\bar{m}_2}(1)\bar{\gamma }^{j_3-\bar{m}_3}(\infty)\prod_{t=1}^{s_\sigma}\bar{\beta}(\bar{w_t})^{(k-2)}\right \rangle _{S_0}  \label{hsd} 
\end{eqnarray}
where, invoking projective invariance, we have fixed the insertions as $(z_1,z_2,z_3)=(0,1,\infty)$. The tilde over $\phi $ refers to the fluctuations $\tilde{\phi}=\phi-\phi_0$ around the zero-mode $\phi_0$. The subindex in ${S_0}$ refers to the fact that the expectation values are defined by the action (\ref{SWZNW}) with $\mu=0$. For simplicity, we have chosen $j_1=m_1=\bar{m}_1$. This is merely to simplify the combinatorics when Wick contracting the ghost fields. The factor $\mu^{s_{\sigma }}\Gamma(-s_{\sigma })$ in (\ref{hsd}) arises through the integration over the zero mode $\phi_0 $ \cite{Goulian}.

Let us briefly comment on the spacetime interpretation of a given amount of screening operators in a correlator that is meant to describe a string scattering amplitude. This has been lucidly discussed in reference \cite{dFK} by di Francesco and Kutasov for the case of string theory in 1+1 dimensions, which involves a spacelike Liouville part. Also there, the $n$-point amplitudes exhibit a prefactor $\Gamma(-s)$, with $s$ being the amount of screening operators to be inserted when realizing the worldsheet correlators in the Coulomb gas approach. Such factor has exactly the same mathematical origin as the one we obtain here in (\ref{hsd}), i.e. it appears through the zero-mode of the field $\phi $. For $s>0$ the Liouville amplitudes are dominated by the region $\varphi \to -\infty$ in the zero-mode integral, which is in the region far from the Liouville wall. In contrast, those with $s<0$ receive their main contribution from the region where the presence of the wall is felt. From the spacetime point of view, amplitudes corresponding to values $s\in \mathbb{Z}_{>0}$ represent scattering processes that take place in the bulk and, as such, are not very sensitive to the details of the wall. Such processes can be interpreted as resonances with the Liouville wall tachyonic composites. A similar interpretation holds in the WZW theory and, likely, it also applies in the case of the timelike theory, although the spacetime picture in the latter is substantially more elusive. In any case, the $\Gamma(-s)$ prefactor in (\ref{hsd}) has an analogous origin and one can think of the correlators with kinematic configurations such as $s<0$ as representing processes that take place in the bulk when the tachyon condensate is not dominant.

Computing the Wick contractions using the free field propagators (\ref{propagators}), one finds \cite{Becker}
\begin{eqnarray}\nonumber
\mathcal{A}_{j_1,m_2,m_3}^{j_1,j_2,j_3}&=&\mu^{s_\sigma}\Gamma(-s_{\sigma})\int \prod_{t=1}^{s_{\sigma }}d^{2}w_{t}\left | w_{t} \right |^{4 j_{1}}\left | 1-w_{t} \right |^{4 j_{2}}\prod_{t<r}\left | w_{t} -w_{r}\right |^{-2\sigma \alpha_{+}^{2}}\\ 
&&\times \lim_{w_{t}^{(n)}\rightarrow w_{t}^{(1)}=w_{t}} \mathcal{P}^{-1} \frac{\partial^{(k-2)s_\sigma} \mathcal{P}}{\partial w_1^{(1)}...\partial w_1 ^{(k-2)}...\partial w_{s_{\sigma}}^{(1)}...\partial w_{s_{\sigma}}^{(k-2)}}\times \ c.c. \  ,
\end{eqnarray}
where
\begin{eqnarray}
\mathcal{P}=\prod_{t=1}^{s_\sigma}\prod_{n=1}^{(k-2)}(1-w_t^{(n)})^{m_2-j_2}\prod_{t<r}^{s_\sigma} \prod_{p=1}^{(k-2)}\prod_{q=1}^{(k-2)} (w_t^{(p)}-w_r^{(q)}),
\end{eqnarray}
and where $c.c.$ stands for its complex conjugate counterpart\footnote{To be precise, this is not exactly the complex conjugate part as it involves $\bar{m}$ instead of $m$.}
\begin{eqnarray}
\bar{\mathcal{P}}=\prod_{t=1}^{s_\sigma}\prod_{n=1}^{(k-2)}(1-\bar{w}_t^{(n)})^{\bar{m}_2-j_2}\prod_{t<r}^{s_\sigma} \prod_{p=1}^{(k-2)}\prod_{q=1}^{(k-2)}(\bar{w}_t^{(p)}-\bar{w}_r^{(q)}).
\end{eqnarray}

This yields
\begin{eqnarray}
\mathcal{P}^{-1}  \frac{\partial^{(k-2)s_\sigma} \mathcal{P}}{\partial w_1^{(1)}...\partial w_1 ^{(k-2)}...\partial w_{s_{\sigma}}^{(1)}...\partial w_{s_{\sigma}}^{(k-2)}}=\frac{\Gamma(-j_2+m_2+(k-2)s_{\sigma})}{\Gamma({m}_2-j_2)}\prod_{t=1}^{s_\sigma}(1-w_t)^{-(k-2)}.
\end{eqnarray}

Putting all together, and assuming for simplicity that $m_{2,3}=\bar{m}_{2,3}$, one finds
\begin{eqnarray}\nonumber \label{structureconstant}
\mathcal{A}_{j_1,m_2,m_3}^{j_1,j_2,j_3}&=&\mu^{s_\sigma}\Gamma(-s_{\sigma})(-1)^{\sigma s_\sigma{\alpha_{+}^{2}}/{2}}\gamma(j_2-m_2+1)\gamma(j_3-m_3+1)\\ 
&&\times\int \prod_{t=1}^{s_{\sigma }}d^{2}w_{t}\left | w_{t} \right |^{4j_{1}} \left | 1-w_{t} \right |^{4 j_2 -\sigma\alpha_{+}^{2}}\prod_{t<r}\left | w_{t} -w_{r}\right |^{-2\sigma \alpha_{+}^{2}},
\end{eqnarray}
where we have defined\footnote{Do not mistake this function for the ghost fields $\gamma(z)$.} $\gamma(x)={\Gamma(x)}/{\Gamma(1-x)}.$ See \cite{Becker} and \cite{GN3} for details.

The integral in (\ref{structureconstant}) can be performed by using the Fateev-Dotsenko formula \cite{Dotsenko} (see formulas (B.9) and (B.10) therein). The result reads
\begin{eqnarray}\nonumber 
&&\mathcal{A}_{j_1,m_2,m_3}^{j_1,j_2,j_3}=\mu^{s_\sigma}\pi^{s _\sigma}\Gamma(-s_{\sigma})\Gamma(1+s_\sigma )(-1)^{\sigma s_\sigma {\alpha_{+}^{2}}/{2}}\gamma(j_2-m_2+1)\gamma(j_3-m_3+1) \\ 
&& \times \left( \gamma (-\sigma {\alpha_{+}^{2}}/{2} ) \right )^{-s_\sigma} \prod_{t=0}^{s_\sigma-1} \frac{\gamma (-(t+1)\sigma \alpha_+^2/2)\gamma (-1-2 j_1 -2 j_2 +(s_\sigma +t)\sigma {\alpha_{+}^{2}}/{2})}{\gamma (-2 j_1 +t\sigma {\alpha_{+}^{2}}/{2} )\gamma (-2 j_2+(1+t)\sigma{\alpha_{+}^{2}}/{2})}   \label{Uh}
\end{eqnarray}


Let us now recall how to compute this $3$-point function in the standard case $k>2$. We do this with the purpose of spotlighting the steps in the analytic extension in $s$, which is the key point to understand the difference with respect to the subcritical case $k<2$. 

When $k>2$, we have to calculate (\ref{Uh}) for the case $\sigma=+ 1$. The strategy is, as said, starting with the assumption $s_+ = {2} ( 1+j_2 + j_2 + j_3)/\alpha^2_+ \in \mathbb{Z}_{> 0}$ and, then, analytically extend the result to complex values $s_+\in \mathbb{C}$ by a controlled systematic procedure. The assumption $s_+\in \mathbb{Z}_{> 0}$ is of course necessary to make sense out of the products appearing in (\ref{Uh}). This is at the root of the difference with the case $k<2$, where we have to deal with expressions with $s_- = -s_+$. More concretely, when going from $k>2$ to $k<2$ the products that appear in the Coulomb gas realization transform as follows
\begin{equation}
\prod_{i=1}^{s_+} f(i) \rightarrow \prod_{i=1}^{-s_+} f(i) \ . \label{cacatua18}
\end{equation} 

The analytic extension of (\ref{Uh}) to other values of $s_+$ demands to take care of the different factors appearing in that expression, one by one. A first simple observation is that, assuming $s_+\in \mathbb{Z}_{> 0}$, 
\begin{eqnarray}
\prod_{t=0}^{s_{+}-1}\gamma(1+2j_1-t{\alpha_{+}^2}/{2})&=&\gamma(1+2j_1)\prod_{t=1}^{s_{+}-1}\gamma^{-1}(-2j_1+t{\alpha_{+}^2}/{2}),
\end{eqnarray}
and
\begin{eqnarray}
\prod_{t=0}^{s_{+}-1}\gamma (1+2 j_2-(t+1){\alpha_{+}^{2}}/{2})&=&\gamma(j_2-j_1-j_3)\prod_{t=1}^{s_{+}-1}\gamma^{-1}(-2j_2+t{\alpha_{+}^2}/{2}).
\end{eqnarray}

Next, using basic properties of the $\Gamma$-function and the definition of the function $G_k$ given in Appendix A, one can write
\begin{eqnarray} \label{produc1+}
\prod_{t=1}^{s_{+}} \gamma \left(-t{\alpha_{+}^2}/{2} \right)=
\lim_{\epsilon\to 0}\frac{1}{\Gamma(\epsilon)}\gamma (-1-j_1-j_2-j_3)\frac{G_k(-j_1-j_2-j_3-2)}{G_k(-1)}\left(\frac{\alpha_{+}^2}{2}\right)^{\eta _-}.
\end{eqnarray}
with ${\eta _-}={-s_+-s_+(s_+-1){\alpha_{+}^2}/{2}}$. The divergent factor $\Gamma (0)$ appearing in the denominator eventually cancels out with other contribution when assembling the different pieces. 

One also finds
\begin{eqnarray} \label{product2+}
\prod_{t=1}^{s_{+}-1} \gamma (-j_1-j_2+j_3+t{\alpha_{+}^2}/{2}) \prod_{i=1}^{2} \gamma (1+2j_i-t{\alpha_{+}^2}/{2}) = \prod_{i=1}^{3}\frac{G_k(2j_i-j_1-j_2-j_3-1)}{G_k(-1-2j_i)} \left(\frac{\alpha_{+}^2}{2}\right)^{\eta_+} 
\end{eqnarray}
with ${\eta_+}={1-s_+ +s_+ (s_+ -1){\alpha_{+}^2}/{2}}$.
 
Putting all together and considering other properties of the $\Gamma $-function, in particular that for $n\in \mathbb{Z}_{\geq 0}$ one has
\begin{equation}
\lim_{\epsilon \to 0} \frac{\Gamma(1-n+\epsilon )\Gamma(n-\epsilon ) }{\Gamma(-\epsilon)} = (-1)^n , \label{La23}
\end{equation}
one finally obtains \cite{GN3}
\begin{eqnarray} 
\mathcal{A}_{j_1,m_2,m_3}^{j_1,j_2,j_3}&=&\mu^{s_+}\pi^{s _{+}}(-1)^{s_+ {\alpha_{+}^2}/{2} }\left({\alpha_{+}^2}/{2}\right)\left( \gamma ( {\alpha_{+}^2}/{2}) \right)^{s_{+}} \gamma(j_2-m_2+1)\gamma(j_3-m_3+1)  \nonumber \\
&&\gamma(-1-j_1-j_2-j_3)\gamma(1+2j_1)\gamma(-j_1-j_2+j_3)\gamma(j_2-j_1-j_3)  \nonumber \\ 
&& \frac{G_k(-2-j_1-j_2-j_3)}{G_k(-1)}\prod_{i=1}^{3} \frac{G_k(2j_i-j_1-j_2-j_3-1)}{G_k(-1-2j_i)}, \label{finalfinalD+}
\end{eqnarray}
which is the $k>2$ $H_3^+$ WZW 3-point function involving a highest-weight state of the $SL(2,\mathbb{R})$ representation in the so-called $m$-basis.

It is convenient to write expression (\ref{finalfinalD+}) in terms the function $\Upsilon _b$ which is usually employed to write the Dorn-Otto-Zamolodchikov-Zamolodchikov (DOZZ) formula of Liouville theory \cite{DO, ZZ}. This is achieved by considering the relation
\begin{equation}\label{relation}
G_k(x)=b^{-b^2x^2-(b^2 +1)x}\Upsilon_b^{-1}(-bx),
\end{equation}
with $b^{-2}=k-2$; see Appendix A. In terms of this function, the result reads
\begin{eqnarray}
\mathcal{A}_{j_1,m_2,m_3}^{j_1,j_2,j_3}=(-1)^{b^{-2} s_+}\gamma(1+j_2-m_2)\gamma(1+j_3-m_3) \ {\mathcal{I}}(j_1,j_2,j_3,b^{-2}). \label{44}
\end{eqnarray}
\begin{eqnarray}
{\mathcal{I}}(j_1,j_2,j_3,b^{-2})=\mu^{s_+}\pi^{s _{+}}b^{-4}\left ( \gamma ( b^{-2}) \right )^{s_{+}}\mathcal{D}_b(j_1,j_2,j_3).
\end{eqnarray}
\begin{eqnarray}
\mathcal{D}_b(j_1,j_2,j_3)=\gamma(-1-j_1-j_2-j_3)\gamma(1+2j_1)\gamma(j_2-j_1-j_3)\gamma(j_3-j_1-j_2)C_b(j_1,j_2,j_3),
\end{eqnarray}
\begin{eqnarray}
C_b(j_1,j_2,j_3)=\frac{b^{-2b^2(\sum_{i=1}^3 j_i+1)+3}\Upsilon_{b}(b)}{\Upsilon_{b}(b(2+j_1+j_2+j_3))}
\prod_{i=1}^{3} \frac{\Upsilon_{b}(b(2j_i+1))}{\Upsilon_{b}(b(j_1+j_2+j_3-2j_i+1))}. \label{47}
\end{eqnarray}

This expression for $C_b(j_1,j_2,j_3)$ exactly reproduces the formula for the WZW structure constants obtained in Ref. \cite{Teschner}. To see this explicitly, one has to take Eq. (64) in \cite{Teschner}, consider the Weyl reflection $j\to -1-j$, and perform the Mellin transform from the $x$-basis to the $m$-basis. This exactly reproduces (\ref{finalfinalD+}); see \cite{GN3} for details.

\section{The subcritical theory}

Consider now the continuation of the 3-point function to values $k<2$. As we will see, the answer is far from being obvious. In fact, we will observe here a phenomenon similar to what happens in Liouville field theory, where the spacelike and the timelike 3-point functions are, roughly speaking, one the inverse of the other. 

Let us go back to expression (\ref{Uh}) and consider now $\sigma =-1$. Namely, consider
\begin{eqnarray}\nonumber \label{sigma-1}
\mathcal{A}_{j_1,m_2,m_3}^{j_1,j_2,j_3}&=&\mu^{s_-}\pi^{s _-} \Gamma(1+s_-)\Gamma(-s_{-})(-1)^{- {\alpha_{+}^{2}}s_-/2}\gamma(j_2-m_2+1)\gamma(j_3-m_3+1) \\ 
&&\times \left ( \gamma ({\alpha_{+}^{2}}/{2} ) \right )^{-s_-} \prod_{t=0}^{s_{-}-1} \frac{\gamma \left((t+1){\alpha_{+}^{2}}/{2} \right)\gamma (-1-2 j_1 -2 j_2 -(s_- +t){\alpha_{+}^{2}}/{2})}{\gamma (-2 j_1 -t{\alpha_{+}^{2}}/{2} )\gamma (-2 j_2-(t+1){\alpha_{+}^{2}}/{2})}
, \label{rulito}
\end{eqnarray}
with $s_-=-2(1+j_1+j_2+j_3)/\alpha^2_+$.

The 3-point function (\ref{44})-(\ref{47}) for a $k>2$ theory was obtained by starting from the case $s_+=2(1+j_1+j_2+j_3)/\alpha_+^2\in \mathbb{Z}_{>0}$. In the case $k<2$, taking into account that $s_-=-2(1+j_1+j_2+j_3)/\alpha_+^2=-s_+\in \mathbb{Z}_{<0}$, a natural proposal would be trying to analytically extend the multiple products appearing in expression (\ref{rulito}) to negative integers values of $s_-=-s_+\in\mathbb{Z}_{\leq 0}$. This proposal is similar to the one used in \cite{Giribet1} to solve the timelike Liouville theory, which was shown to reproduce the correct expression \cite{Zamolodchikov}. Then, the question reduces to that of how making sense out of the products in (\ref{rulito}) for negative values of $s_-$. This question is actually not new in the context of CFT. A similar problem appears, for instance, in Minimal Models coupled to 2D gravity. In Ref. \cite{Dotsenko2}, Dotsenko proposed a trick to making sense out of the product of a negative amount of screening charges. This follows from extending the basic property
\begin{eqnarray}
P(l)\equiv \prod_{i=1}^{l}f(i)=\frac{\prod_{i=1}^{\infty}f(i)}{\prod_{i=l+1}^{\infty}f(i)}=\frac{P(\infty )}{\prod_{i=1}^{\infty}f(i+l)}, 
\end{eqnarray}
which is obviously valid for positive integer $l\in \mathbb{Z}_{>0}$, to negative values of $l$ by simply extrapolating the functional properties of $P(l)$. Then, as in \cite{Dotsenko2}, one may consider extending the definition of the function $P(l)$ to values $l=-\left | l \right |$ by defining
\begin{eqnarray}\label{negative}
P(l)\equiv \prod_{i=0}^{\left | l \right |-1}\frac{1}{f(-i)}  \ \ \ \text{for} \ l\in \mathbb{Z}_{<0}.
\end{eqnarray}

This trick, which has been used in many different examples before, can also be applied succesfully to our case. In this way, we can give meaning to the Coulomb gas formulas for the subcritical 3-point correlation function.

Making use of (\ref{negative}) and iterating the shift properties of the $\Gamma $-function and $G_k$-function, we can rewrite each of the factors in (\ref{Uh}) in terms of the latter function. For instance, using 
\begin{equation}
G_k(x-{\alpha_{+}^2}/{2})=\gamma(1+x)G_k(x)\left(\frac{\alpha_+^2}{2}\right)^{-2x-1},
\end{equation} 
the first multiple product in the numerator of the second line of (\ref{Uh}) can be written as
\begin{eqnarray}\label{produc1-}
\prod_{t=0}^{s_{-}-1} \gamma ((t+1){\alpha_{+}^2}/{2} )=\gamma(-1- j_1-j_2-j_3)\frac{G_k(-1)}{G_k(j_1+j_2+j_3+1)}\left(\frac{\alpha_{+}^2}{2}\right)^{\eta_0} 
\end{eqnarray}
with $\eta _0 = (\alpha_+^2/2)(s_-^2 -s_-)-s_- +1$. Analogously, we can write 
\begin{eqnarray}\nonumber \label{product2-}
\prod_{t=0}^{s_{-}-1}\gamma (1+2j_1+t{\alpha_{+}^2}/{2})&=&\frac{\gamma(1+2j_1)G_k(2j_1)}{G_k(-j_1+j_2+j_3)} \left(\frac{\alpha_{+}^2}{2}\right)^{\eta_1}
\end{eqnarray}
with $\eta _1 = s_- + 4j_1 (s_- -1) +(\alpha_+^2/2) (s_-^2 - s_-)-1$. Considering that $\gamma^{-1}(1-j_2+j_1+j_3)=\gamma(j_2-j_1-j_3)$, one can express
\begin{eqnarray}\nonumber \label{product3-}
\prod_{t=0}^{s_{-}-1}\gamma (1+2j_2+(t+1){\alpha_{+}^2}/{2})&=&\frac{\gamma(j_2-j_1-j_3) G_k(2j_2)}{G_k(-j_2+j_1+j_3)}\left(\frac{\alpha_{+}^2}{2}\right)^{\eta_2}
\end{eqnarray}
with $\eta _2 = s_- + 4j_2 (s_- -1)+(\alpha_+^2/2) (s_-^2 - s_-)-1$; and using that $\gamma^{-1}(1-j_3+j_1+j_2)=\gamma(j_3-j_1-j_2)$, one finds
\begin{eqnarray}\nonumber\label{product4-}
\prod_{t=0}^{s_{-}-1}\gamma (-1-2j_1-2j_2-(s_-+t){\alpha_{+}^2}/{2})&=&\frac{\gamma(j_3-j_1-j_2) G_k(2j_3)}{G_k(-j_3+j_1+j_2)}\left(\frac{\alpha_{+}^2}{2}\right)^{\eta_3}
\end{eqnarray}
with $\eta _3 = s_- + 4j_3 (s_- -1) +(\alpha_+^2/2) (s_-^2 - s_-)-1$. Notice that $\eta_0+\eta_1+\eta_2+\eta_3= -2(s_{-}-1)$.

Combining all this, we eventually find
\begin{eqnarray}\nonumber \label{finalD-}
\mathcal{A}_{j_1,m_2,m_3}^{j_1,j_2,j_3}&=&\mu^{s_-}\pi^{s _{-}}\Gamma(-s_{-})\Gamma(1+s_{-})(-1)^{s_-(1-{\alpha_{+}^2}/{2})}\Big({\alpha_{+}^2}/{2}\Big)^{2}\left( \gamma (-{\alpha_{+}^2}/{2} ) \right)^{s_{-}} \\ \nonumber
&&\times\gamma(j_2-m_2+1)\gamma(j_3-m_3+1)\gamma(-1-j_1-j_2-j_3)\gamma(1+2j_1)\gamma(j_2-j_1-j_3) \nonumber \\
&&\times\gamma(j_3-j_1-j_2)\frac{G_k(-1)}{G_k(j_1+j_2+j_3+1)} \prod_{i=1}^{3} \frac{G_k(2j_i)}{G_k(j_1+j_2+j_3-2j_i)}, \label{Bardo}
\end{eqnarray}
which is our main result: This is the 3-point correlation function for the subcritical WZW theory.

As we did for the case $k>2$, we can easily express (\ref{Bardo}) in terms of the $\Upsilon_b$-functions. Denoting $\hat{b}=ib$, which means $\hat{s}_-=-\hat{b}^2( j_1+j_2+j_3+1)$, this yields
\begin{eqnarray}
\mathcal{\hat{A}}_{j_1,m_2,m_3}^{j_1,j_2,j_3}=(-1)^{(-\hat{b}^{-2})\hat{s}_-}\gamma(1+j_2-m_2)\gamma(1+j_3-m_3)\hat{{\mathcal{I}}}(j_1,j_2,j_3,-\hat{b}^{-2}). \label{67}
\end{eqnarray}
\begin{eqnarray}
\hat{{\mathcal{I}}}(j_1,j_2,j_3,-\hat{b}^{-2})=\mu^{\hat{s}_-}\pi^{\hat{s} _{-}}\hat{b}^{-4}\left ( \gamma (-\hat{b}^{-2} ) \right )^{\hat{s}_{-}}\hat{\mathcal{D}}_{\hat{b}}(j_1,j_2,j_3).
\end{eqnarray}
\begin{eqnarray}
\hat{\mathcal{D}}_{\hat{b}}(j_1,j_2,j_3)=\gamma(-1-j_1-j_2-j_3)\gamma(1+2j_1)\gamma(j_2-j_1-j_3)\gamma(j_3-j_1-j_2)\hat{C}_{\hat{b}}(j_1,j_2,j_3),
\end{eqnarray}
\begin{eqnarray} \label{C-timelike}
\hat{C}_{\hat{b}}(j_1,j_2,j_3)=\hat{b}^{2+2\hat{b}^2+2\hat{b}^2(\sum_{i=1}^3 j_i)} \frac{\Upsilon _{\hat{b}}(-\hat{b}(j_1+j_2+j_3+1))}{\Upsilon _{\hat{b}}(\hat{b})}\prod_{i=1}^{3}
\frac{\Upsilon _{\hat{b}}(\hat{b}(2j_i-j_1-j_2-j_3))}{\Upsilon _{\hat{b}}(-\hat{b}(2j_i))},
\end{eqnarray}
where we have used (\ref{La23}) and have excluded the divergent factor $\Gamma(0)$. The same type of divergence appears in timelike Liouville field theory \cite{Giribet1}.

The remarkable fact is that (\ref{C-timelike}) is, roughly speaking, the inverse of (\ref{47}) and not the simple extension $b\to \hat{b}$ that one could have naively expected. In fact, one observes the remarkable factorization\footnote{The $b$-dependent contribution on the right hand side is irrelevant, as it can be absorbed in the normalization of the vertices.}
\begin{equation} \label{Ufsilaposta}
C_b(j_1, j_2, j_3) \hat{C}_b(-1-j_1, -1-j_2, -1-j_3) = b^{5-6b^2-4b^2\sum_{i=1}^3 j_i } .
\end{equation}

This type of inversion phenomenon had already been observed by Zamolodchikov in the case of Liouville theory coupled to Generalized Minimal Models \cite{Zamolodchikov}, where the 3-point functions also factorize in a similar way. This is also behind the surprising cancellation that superstring amplitudes\footnote{As explained in \cite{Nicolas}, supersymmetry is crucial for this simplification to occur in string theory on AdS$_3\times S^3 \times T^4$ NS-NS backgrounds.} in AdS$_3 \times S^3$ backgrounds exhibit \cite{Nicolas, Pakman1, Pakman2}.

Expression (\ref{Ufsilaposta}) manifestly shows the non-trivial form that the 3-point function takes when analytically extending from $b\in \mathbb{R}$ to $ib$. At classical level, the extension to imaginary values of $b$ is straightforward, but this is not the case at quantum level where the problem of defining the correct integration cycle in the path integral is subtle \cite{H-M-Witten}. In contrast to the 3-point function, the 2-point function does admit a straightforward extension from real to imaginary values of $b$. Consequently, this raises the question as to how the 2-point function can be recovered from the timelike structure constants (\ref{C-timelike}) in the appropriate limit. In the case $b\in \mathbb{R}$ this smoothly follows from the functional property
\begin{equation}
\lim_{\varepsilon \to 0 } \frac{G_k(j_2-j_3+\varepsilon )G_k(j_3-j_2+\varepsilon )}{G_k(-1)G_k(1-2\varepsilon )} = 2\pi i \ \frac{\Gamma\Big(1+\frac{1}{k-2}\Big)}{\Gamma\Big(1-\frac{1}{k-2}\Big)}
\ \delta(j_2-j_3)
\end{equation}
which leads to reobtain the reflection coefficient\footnote{To be precise, the physical interpretation of the 2-point function in the timelike case is not that of a reflection coefficient, but that of the quantity whose modulus gives the particle production rate in the time-dependent background \cite{StromingerTakayanagi, Yasuaki2}.} $\mathcal{A}_{m_2,m_3}^{j_2,j_3}\sim B(j_2)\delta (j_2-j_3)$ as the limit of the structure constant $\lim _{j_1\to 0} C(j_1,j_2,j_3)$. In the timelike theory, in contrast, and due to the special dependence of the $G_k$-functions in (\ref{Bardo}), the relation between the 2- and the 3-point function is different and somehow arbitrary, exactly as it occurs in the Liouville theory with $c<1$. In particular, this arbitrariness manifests itself in that the timelike Liouville 3-point function evaluated on momenta $\alpha_1=0$, $\alpha_2\neq 0\neq \alpha_3$ does not develop a delta function $\sim\delta (\alpha_2 - \alpha_3) $ (see Appendix B). This peculiar feature has been discussed, for instance, in reference \cite{H-M-Witten} (see discussion in section 7.1 therein), and this had been previously discussed in references \cite{StromingerTakayanagi} and \cite{McElgin}. In \cite{McElgin}, this feature was taken as evidence that in the $c<1$ Liouville theory the 2-point function is genuinely non-diagonal in the conformal dimension. This raises the question as to whether or not the theory should be regarded as an actual CFT. In \cite{H-M-Witten} a possible interpretation for this phenomenon was given: It was suggested there that the limit $\alpha_1 \to 0$ of the 3-point function should probably not be interpreted as the limit in which one of the vertex operators tends to the identity operator, but rather as the limit in which an alternative dimension-zero operator emerges in the correlator. For a non-unitary CFT this is certainly a possibility, and the same interpretation is possible in our sine-Liouville computation. On general grounds, and even when we are unable to give a final resolution of this problem, we certainly expect that the explanation to this peculiar feature in sine-Liouville theory will be the same as in Liouville theory. Notice that in sine-Liouville theory, as well as in its FZZ dual, there are natural candidates for such non-trivial dimension-zero operator, given by the conjugate identities $j=\pm m=\pm \bar{m}=-k/2$ that, according to Fateev, Zamolodchikov, and Zamolodchikov are essential elements to construct the spectrum of the theory \cite{FZZ}. In any case, the non-diagonal form that the 2-point function takes in in zero-momentum limit of the 3-point function requires further investigation. 

\section{Modular functions} 

With the intention of comparing our result (\ref{C-timelike}) with what happens in the process of analytically extending Liouville theory to values $c<1$, let us express the $3$-point function (\ref{C-timelike}) in terms of the function 
\begin{equation}
H_b(x) = \Upsilon_b(x)\Upsilon_{ib}(-ix+ib), \label{YUI}
\end{equation}
introduced in \cite{Zamolodchikov}. This will allow us to write, as it happens in Liouville theory, the quotient of the timelike and spacelike formulas in terms of Jacobi functions.

Given the relation $\hat{b}=ib$, one finds
\begin{eqnarray}\nonumber
\hat{C}_{\hat{b}}(j_1,j_2,j_3)&=& e^{i\pi (1-b^2 -b^2 \sum_{i=1}^3j_i )}b^{2-2b^2-2b^2(\sum_{i=1}^3 j_i)}\frac{H _b(b(\sum_i j_i+2))}{H _b(0)}
\prod_{i=1}^{3}
\frac{H _b(b(-2j_i+j_1+j_2+j_3+1))}{H _b(b(2j_i+1))}\\ 
&&\times
\frac{\Upsilon_{b}(0)}{\Upsilon_{b}(b(2+j_1+j_2+j_3))}
\prod_{i=1}^{3}\frac{\Upsilon_{b}(b(2j_i+1))}{\Upsilon_{b}(b(-2j_i+j_1+j_2+j_3+1))}
\end{eqnarray}
where we have reinserted the divergent factor $\Gamma(0)$ that combines with a $\Upsilon_b(b)$ factor and the number $\Upsilon_b(0)$ that appear in the calculation. Notice that $\Upsilon_b(b)= \Gamma(0)\Upsilon_b(0)$ and $\Upsilon_b(0)\Upsilon_{\hat{b}}(\hat{b})=H_b(0)$. 

On the other hand, the following relation holds
\begin{equation}
H_b(x) = e^{\frac{i\pi }{2}(x^2+xb^{-1}-xb+b^2/4-3b^{-2}/4-1/4)} \frac{\theta_1 (xb^{-1},b^{-2})}{\theta_1 (1/2+b^{-2}/2,b^{-2})}, \label{TYUY}
\end{equation}
where $\theta_1$ is the Jacobi function, whose definition can be found in the Appendix A. Comparing this with the formula for $C_b(j_1,j_2,j_3)$, we find
\begin{eqnarray}
\frac{\hat{C}_{ib}(j_1,j_2,j_3)}{C_b(j_1,j_2,j_3)}=  \frac{\theta_1\left(j_1+j_2+j_3+1,b^{-2}\right)}{b\ \theta_ 1(-1,b^{-2})}
\prod_{i=1}^{3}
\frac{\theta_1\left(-2j_i+j_1+j_2+j_3,b^{-2}\right)}{\theta_1\left(2j_i,b^{-2}\right)} \label{Cuarenta}
\end{eqnarray}
where we have used the modular properties of the Jacobi function, namely $\theta_1 (x+1,\tau ) = e^{-i\pi }\theta_1 (x,\tau )$. 

We observe that expression (\ref{Cuarenta}) is analogous to Eq. (7.42) of Ref. \cite{H-M-Witten} for the timelike Liouville theory (see Appendix B). 

\section{Other proposals}

Before concluding, let us comment about the comparison of our formula for $\hat{C}_{\hat{b}}(j_1,j_2,j_3)$ with other proposals made in the literature for the subcritical WZW 3-point function. In \cite{Yasuaki2}, Hikida and Takayanagi gave such a formula by extrapolating the expressions obtained by Schomerus in \cite{Schomerus} for the $c<1$ CFT. The expression of \cite{Yasuaki2}, however, has been written in terms of the $H_b$- and $Y_{\beta }$-functions introduced in \cite{Schomerus} and, thus, a little of extra work is necessary to compare with (\ref{C-timelike}) (see Appendix A). An efficient way of verifying the consistency between our result (\ref{Bardo}) and the formula for the 3-point function of \cite{Yasuaki2} is to compute for the latter the quotient between the $k>2$ and the $k<2$ cases and compare this with (\ref{Cuarenta}). One might start from the 3-point function for $k>2$, which corresponds to expression (4.15) of reference \cite{Yasuaki2}. Performing $b\to \hat{b}=ib$, and considering (\ref{TYUY}) that relates the $H_b$-function with the Jacobi $\theta _1$-function, one can write the ratio of the 3-point functions as a quotient of $\theta_1$ functions. In order to compare with \cite{Yasuaki2} it is necessary to take into account the dictionary between their parameters and ours; namely, to consider
\begin{equation}
b=-\frac{i}{\sqrt{\alpha }} \ , \ \ \ \ j_i = -1+i\sqrt{\alpha }\omega_i. 
\end{equation}
It is also convenient to shift the zero mode $\phi_0$ in order to set $\mu = {\Gamma(1-b^2)}/(\pi {\Gamma(1+b^2))}$, which set $\nu (k) $ of \cite{Yasuaki2} to 1. These definitions lead to the expression (4.17) in \cite{Yasuaki2}, namely\footnote{It is also needed to take into account an extra factor 1/2 in the convention used in Eq. (4.16) of \cite{Yasuaki2} when writing the function $H$.}
\begin{equation}
H(w)=\theta_1(j,\alpha)\ Y_{\frac{1}{\sqrt{\alpha}}}(w)
\end{equation}
where the function $Y_{\beta}$, which is defined in Appendix A, satisfies the relation
\begin{equation}
Y_{\beta}(w)=\Upsilon_{\beta}^{-1}(iw +\beta).
\end{equation}

Considering these functional relations, a careful examination of the expressions revels that our result (\ref{Bardo}) relates to the Hikida-Takayanagi result for the $k<2$ WZW 3-point function \cite{Yasuaki2} in the same manner as how the Harlow-Maltz-Witten result for Liouville timelike 3-point function \cite{H-M-Witten} relates to the formula proposed by Schomerus in \cite{Schomerus} for the $c<1$ theory. Actually, this is not a surprise since the expression proposed in \cite{Yasuaki2} has been constructed by following the procedure described in \cite{Schomerus}.

\section{Conclusions}

In this paper we have revisited the problem of non-homogeneous tachyon condensation in bosonic closed string theory. A worldsheet theory describing this phenomenon is given by the Lorentzian continuation of the Sine-Liouville model. We solved this model exactly, in the sense of having provided a closed expression of the 3-point correlation functions on the Riemann sphere at finite $\alpha '$. The strategy was resorting to the T-dual description in terms of the gauged $H_+^3 / U(1)$ Wess-Zumino-Witten model at subcritical level $k<2$. Using the same Coulomb gas techniques which in the case of timelike Liouville field theory have shown to reproduce the correct 3-point function, we derived here a formula for the WZW correlation functions within the range $0<k<2$. This result is given by expression (\ref{Bardo}), which is our main result; see also (\ref{67})-(\ref{C-timelike}). This represents an exact solution to string theory on a time-dependent background.  

Remarkably, our result for the subcritical WZW 3-point function turns out to be, roughly speaking, the inverse of the standard (i.e. $k>2$) result and not the simple extension $\sqrt{k-2}\to i\sqrt{k-2}$ of it that one could have naively expected. This phenomenon is, mutatis mutandis, the same that what happens in timelike Liouville theory, and which had already been noticed by Al. Zamolodchikov in the context of Generalized Minimal Models. In other words, our expression (\ref{C-timelike}) relates to the standard WZW 3-point function in a similar way as how the Harlow-Maltz-Witten $3$-point function of timelike Liouville theory relates to the analytic continuation of the Dorn-Otto-Zamolodchikov-Zamolodchikov structure constants. On the one hand, the ratio between both correlators admits a simple expression as a quotient of Jacobi's $\theta _1$-functions. Their product, on the other hand, exhibits a remarkable factorization. 

\[
\]

The work of G.G. has been funded by CONICET and UBA through grants PIP 0595/13, UBACyT 20020120100154BA, and NSF-CONICET bilateral agreement. He thanks Sylvain Ribault for interesting remarks on timelike Liouville theory. L.R. thanks CAPES for financial support, and thanks Renato S\'{a}nchez and Elcio Abdalla for valuable discussions.

\section*{Appendix A: Special functions}

\subsubsection*{The $G_k$-function}

The Barnes' double $\Gamma$-function is given by
\begin{equation}
\log \Gamma_2 (x|1,y) = \lim_{\epsilon\to 0} \partial_{\epsilon } \left( \sum_{n,m\in Z_{\geq 0}} (x+n+my)^{-\epsilon } - \sum_{n,m\in Z_{\geq 0} }' (n+my)^{-\epsilon } \right) , \label{prime}
\end{equation}
where the prime on the second sum means that the step $(m,n)\neq (0,0)$ is omitted.

The function $G_k$ is defined in terms of $\Gamma_2$ as follows
\begin{equation}\label{G_k}
G_k(x)\equiv (k-2)^{\frac{x(k-1+x)}{2(k-2)}} \Gamma_2 (-x|1,k-2) \Gamma_2 (k-1+x|1,k-2) .
\end{equation}

This function obeys the reflection relation
\begin{eqnarray}
G_k(x)&=&G_k(-x-k+1) , \label{89}
\end{eqnarray}
and the translation (shift) relations
\begin{eqnarray}
G_k(x+1)&=&\gamma(-(x+1)/(k-2))G_k(x) \label{90} \\
G_k(x-k+2)&=&(k-2)^{-2x-1}\gamma(x+1)G_k(x) , \label{91}
\end{eqnarray}
where the function $\gamma $ is defined in terms of the $\Gamma $-function as follows
\begin{equation}
\gamma(x) = \frac{\Gamma(x)}{\Gamma(1-x)},
\end{equation}
and, thus, obeys
\begin{equation}
\gamma(x)\gamma(1-x)=1, \ \ \ \ \ \gamma(x)\gamma(-x)=-x^{-2},
\end{equation}
because of the property $\Gamma(1+x)=x\Gamma(x)$.

\subsubsection*{The $\Upsilon_k$-function}

Function $G_k$- relates to the $\Upsilon_b$-function as follows
\begin{equation}
\Upsilon_b(-bx)=b^{-b^2x^2-(b^2 +1)x}G_k^{-1}(x),
\end{equation}
with $b^{-2}=k-2$. Therefore, from (\ref{G_k}) it follows that
\begin{equation}\label{Upsilon_b}
 \Upsilon_b(x) = \Gamma_2^{-1} (x|b,b^{-1}) \Gamma_2^{-1} (b+b^{-1}-x|b,b^{-1}).
\end{equation}

Function $\Upsilon_b$ can also be defined by
\begin{equation}
\log \Upsilon_b (x) \equiv \eta_b + \int_{\mathbb{R}_{>0}} \frac{dt}{t} \left( (b+b^{-1} -2x)^2 \frac{e^{-t}}{4} -\frac{\sinh^2((b/2+b^{-1}/2-x)t/2)}{\sinh(bt/2)\sinh(b^{-1}t/2)} \right) . \label{96}
\end{equation}
with the constant $\eta_b = -2\log \Gamma_2 (\frac{b+b^{-1}}{2}|b,b^{-1})$. It obeys the following shift relations
\begin{eqnarray}
\Upsilon_b(x+b) &=& \gamma(bx) b^{1-2bx}\Upsilon_b (x) \label{97} \\
\Upsilon_b(x+b^{-1}) &=& \gamma(b^{-1}x) b^{-1+2b^{-1}x}\Upsilon_b (x) \label{98}
\end{eqnarray}
together with the reflection relations
\begin{eqnarray}
\Upsilon_b(x) &=& \Upsilon_{b^{-1}} (x) \\
\Upsilon_b(x) &=& \Upsilon_b (b+b^{-1}-x), 
\end{eqnarray}
which follow from (\ref{89})-(\ref{91}) and the definition (\ref{96}).

A useful property of $\Upsilon_b$-function, which comes from iterating (\ref{97})-(\ref{98}), is the following
\begin{equation}
\prod_{r=1}^{x} \gamma(rb^2) = \frac{\Upsilon_b (xb+b)}{\Upsilon_b (b)} b^{x(b^2(x+1)-1)},
\end{equation}
provided $x\in \mathbb{Z}_{>0}$.

Function $\Upsilon_b$ has poles at $x=mb+nb^{-1}$ with $m,n\in \mathbb{Z}_{>0}$ and $m,n\in \mathbb{Z}_{\leq 0}$.

\subsubsection*{The $Y_{\beta}$-function}

Interested in the case of imaginary values of $b=i\beta $ (i.e. $\beta \in \mathbb{R}$, which is also a convention frequently employed in the CFT literature to express these functions), one finds convenient to define the function
\begin{eqnarray}\label{Y_b}
Y_\beta (x)\equiv \Gamma_2(\beta +i x\mid \beta,\beta^{-1})\Gamma_2(\beta^{-1}-ix\mid \beta,\beta^{-1}) ,
\end{eqnarray}
which also admits the definition
\begin{equation}
\log Y_{\beta} (x) = h_{\beta } + \int_{\mathbb{R}_{>0}} \frac{dt}{t} \left( ({b+b^{-1}} -2x)^2 \frac{e^{-t}}{4} -\frac{\sin^2((b/2+b^{-2}/2-x)t/2)}{\sinh(\beta t/2)\sinh(\beta^{-1}t/2)} \right) ,
\end{equation}
with $h_{\beta }$ being a $\beta $-dependent constant. This function obeys
\begin{eqnarray}
Y_{\beta} (x+i\beta) &=& \beta^{1-2ix\beta }Y_{\beta } (x) \gamma(ix\beta) \\
Y_{\beta} (x-i\beta^{-1}) &=& \beta^{-1-2ix/\beta }Y_{\beta } (x) \gamma(-ix\beta^{-1}).
\end{eqnarray}

\subsubsection*{The $H_b$- and $\theta_1 $-functions}

Define the function
\begin{equation}
H_b(x) \equiv \Upsilon_b(x)\Upsilon_{ib}(-ix+ib),
\end{equation}
which obeys the relations
\begin{eqnarray}
H_b(x+b) &=& e^{\frac{i\pi}{2}(2bx-1)} H_b(x)  \label{108} \\
H_b(x+b^{-1}) &=& e^{\frac{i\pi}{2}(1-2b^{-1}x)} H_b(x) . \label{109}
\end{eqnarray}

This function can be written in terms of the Jacobi's $\theta_1$-function, namely
\begin{equation}\label{definitionH}
H_b(x) = e^{\frac{i\pi }{2}(x^2+xb^{-1}-xb+b^2/4-3b^{-2}/4-1/4)} \frac{\theta_1 (xb^{-1},b^{-2})}{\theta_1 (1/2+b^{-2}/2,b^{-2})},
\end{equation}
with
\begin{equation}\label{Theta_1}
\theta_1 (x,\tau ) = i\sum_{n\in \mathbb{Z}} (-1)^n e^{i\pi \tau (n-1/2)^2+2\pi i x(n-1/2)}
\end{equation}
with $\text{Im}(\tau )>0$. Properties (\ref{108})-(\ref{109}) follow from the well-known modular properties of the Jacobi's function, namely
\begin{eqnarray}\label{Theta+1}
\theta_1 (x+1,\tau ) &=& e^{-i\pi }\theta_1 (x,\tau ) , \\
\theta_1 (x+\tau ,\tau ) &=& e^{i\pi (1-\tau-2x)}\theta_1 (x,\tau ) 
\end{eqnarray}

\section*{Appendix B: The timelike Liouville theory}

Consider the DOZZ Formula for the standard ($c>25$) Liouville theory
\begin{eqnarray}\nonumber \label{spacelike Liouville}
C(\alpha_1,\alpha_2,\alpha_3)=\left [ \pi \mu \gamma (b^{2})b^{2-2b^{2}} \right ]^{(Q-\sum _{i}\alpha _{i})/b} 
\frac{\Upsilon_0}{\Upsilon_b(\alpha_1+\alpha_2+\alpha_3-Q)}
\prod_{i=1}^{3}
\frac{\Upsilon_b(2\alpha_i)}{\Upsilon_b(\alpha_1+\alpha_2+\alpha_3-2\alpha_i)}
,
\end{eqnarray}
where $Q=b+b^{-1}$, $b\in \mathbb{R}$, and where $\Upsilon_0$ is a constant. 


The timelike DOZZ formula proposed in \cite{H-M-Witten} reads
\begin{eqnarray}\nonumber \label{timelike Liouville}
&&\hat{C}(\hat{\alpha}_1,\hat{\alpha}_2,\hat{\alpha}_3)=\frac{2\pi}{\hat{b}}\left [ -\pi \mu \gamma (-\hat{b}^{2})\hat{b}^{2+2\hat{b}^{2}} \right ]^{(\sum _{i=1}^3\hat{\alpha} _{i}-\hat{Q})/\hat{b}} e^{-i\pi(\sum_{i=1}^3\hat{\alpha}_i-\hat{Q})/\hat{b}}\\ \nonumber
&&\times \frac{\Upsilon_{\hat{b}}(\hat{\alpha}_1+\hat{\alpha}_2+\hat{\alpha}_3-\hat{Q}+\hat{b})}{\Upsilon_{\hat{b}}(\hat{b})}
\prod_{i=1}^{3}
\frac{\Upsilon_{\hat{b}}(\hat{\alpha}_1+\hat{\alpha}_2+\hat{\alpha}_3-2\hat{\alpha}_i+\hat{b})}{\Upsilon_{\hat{b}}(2\hat{\alpha}_i+\hat{b})}
,
\end{eqnarray}
where the parameters are defined as 
\begin{eqnarray}
\hat{\alpha}_i=-i\alpha_i , \ \ \ \hat{b}^2&=&-b^2, \ \ \ \hat{Q}=-iQ.
\end{eqnarray}

Writing (\ref{timelike Liouville}) as a function of the spacelike parameters, one obtains
\begin{eqnarray}
\frac{\hat{C}(\hat{\alpha}_1,\hat{\alpha}_2, \hat{\alpha}_3)}{{C}({\alpha}_1,{\alpha}_2, {\alpha}_3)} = 2\pi i \frac{\theta_1 (b^{-1}(\alpha_1 +\alpha_2 +\alpha_3 ), b^{-2}) }{ \theta_1 ' (0, b^{-2}) }
\prod_{i=1}^{3} \frac{\theta_1 (b^{-1}(\alpha_1 +\alpha_2 +\alpha_3 -2\alpha_i ), b^{-2})}{\theta_1 (2b^{-1}\alpha_i , b^{-2})}
\end{eqnarray}
with $\theta_1 ' (x, b^{-2})= \frac{\partial }{\partial x} \theta_1 (x, b^{-2})$. This is equation (7.40) of \cite{H-M-Witten} and is the analog of (\ref{Cuarenta}) for Liouville field theory.

\end{document}